# Interpretable Cross-Sphere Multiscale Deep Learning Predicts ENSO Skilfully Beyond 2 Years

Rixu Hao[1,2], Yuxin Zhao[1,2*], Shaoqing Zhang[3*], Guihua Wang[4], Xiong Deng[1,2]

[1]College of Intelligent Systems Science and Engineering, Harbin Engineering University; Harbin, 150001, China.
[2]Engineering Research Center of Navigation Instruments, Ministry of Education, Harbin Engineering University; Harbin, 150001, China.
[3]Key Laboratory of Physical Oceanography, Ministry of Education/Institute for Advanced Ocean Study/Frontiers Science Center for Deep Ocean Multispheres and Earth System (DOMES), College of Oceanic and Atmospheric Sciences, Ocean University of China; Qingdao, 266100, China.
[4]Department of Atmospheric and Oceanic Sciences and CMA-FDU Joint Laboratory of Marine Meteorology, Fudan University; Shanghai, 200433, China.

*Corresponding author. Email: zhaoyuxin@hrbeu.edu.cn (Y.Z.); szhang@ouc.edu.cn (S.Z.)

**Abstract:** El Niño-Southern Oscillation (ENSO) exerts global climate and societal impacts, but real-time prediction with lead times beyond one year remains challenging. Dynamical models suffer from large biases and uncertainties, while deep learning struggles with interpretability and multi-scale dynamics. Here, we introduce PTSTnet, an interpretable model that unifies dynamical processes and cross-scale spatiotemporal learning in an innovative neural-network framework with physics-encoding learning. PTSTnet produces interpretable predictions significantly outperforming state-of-the-art benchmarks with lead times beyond 24 months, providing physical insights into error propagation in ocean-atmosphere interactions. PTSTnet learns feature representations with physical consistency from sparse data to tackle inherent multi-scale and multi-physics challenges underlying ocean-atmosphere processes, thereby inherently enhancing long-term prediction skill. Our successful realizations mark substantial steps forward in interpretable insights into innovative neural ocean modelling.

## Introduction

The El Niño–Southern Oscillation (ENSO) represents the main source of interannual variability in the global climate system, and the ability to predict large-scale climate variability and its impacts on global social and environmental systems is highly dependent on the quality of ENSO predictions (*1–5*). With significant advances in ENSO observations and process understanding, considerable progress has been made in associated modelling and prediction in recent decades (*6-10*). Specifically, physics-driven dynamical models remain indispensable for understanding processes and predicting phenomena. However, existing dynamical models generally suffer from systematic biases due to incomplete representations of processes, which hinders realistic modelling and long-term predictions of climate systems (*11–13*). Currently, skilful prediction of ENSO at lead times of more than one year remains challenging (*6*, *14*, *15*).

Recent progress in deep learning and its innovative applications in ocean sciences has offered promising opportunities to enhance the modelling of natural coupled ocean–atmosphere processes (*16–19*). Neural networks, the foundation of deep learning models, are utilized to automatically describe the intrinsic physical relationships from input predictors to output predictands, dispensing with explicit reliance on the physical laws underlying ENSO processes. Such approaches have been proven to significantly increase the modelling accuracy of nonlinear systems, including ENSO predictions (*20–23*). However, owing to the complicated interactions that occur at various spatiotemporal scales and the nonlinear feedback associated with coupled ocean–atmosphere processes, existing approaches fall short in comprehensively describing the cross-sphere and multiscale coupling underlying ENSO (*9*, *10*). This leads to a lack of physical consistency in the model outputs and hinders interpretability, which is necessary for in-depth insights into the underlying mechanisms (*24–26*). Exploratory research in ocean sciences has explored the contributions of ocean drivers with layer-wise relevance propagation (*27*), establishing physical linkages between neural networks and dynamical processes (*28*, *29*). While such efforts provide physical insights, most research has focused on existing interpretation approaches for neural networks (*30*, *31)*, which fail to fully integrate data with physical information, and further limit our understanding of the dynamics and physical robustness underpinning the enhanced deep learning ability. In particular, the cross-sphere and multiscale nature of ENSO complexity remains far from being self-explanatory in modelling and fails to discover intrinsic causality from observations (*26*).

Hybrid models that integrate dynamical processes with deep learning are promising because they leverage the interpretability of dynamical models and the efficient spatiotemporal modelling of deep learning (*19*, *32*). Deep learning components in hybrid models either replace or refine the traditional physical parameterizations of dynamical models (*26*, *33*). To date, physics-guided neural networks as the research frontier for such models have been trained offline to learn parameterizations independently of their interactions with dynamics. Lack of coupling between deep learning and dynamical processes during training may cause significant issues, such as instability and climate drift (*34*). Furthermore, hybrid models have mostly been limited to idealized scenarios (*35*). Thus, hybrid modelling is still under preliminary exploration in ocean sciences, particularly for ENSO prediction, where incorporating physics priors effectively into formalized deep learning modelling remains challenging.

Here, we present PTSTnet, a physics-guided tensor-train spatiotemporal deep learning model for long-term ENSO prediction. It integrates deep learning with cross-scale dynamical processes through an innovative neural network that implements physics-encoding learning for providing model interpretability while reducing data demands. Besides, cross-scale spatiotemporal fusion learning extends prediction lead times substantially. This model enables direct learning of complex dynamic patterns from spatiotemporal observations, successfully predicting periodic complexity and capturing spatiotemporal telecorrelations for lead times of more than 24 months. PTSTnet provides physically interpretable predictions across various ENSO events with prediction skill outperforming existing state-of-the-art dynamical and deep learning models.

## PTSTnet

Considering the inherent complexity, including multivariate ocean–atmosphere interactions, cross-scale coupling, and spatiotemporal teleconnections (*20*, *23*), skilful ENSO prediction requires leveraging both dynamical principles and statistical-learning approaches. PTSTnet (Fig. 1) provides such unification through an innovative neural network that enables end-to-end prediction error optimization, which extends lead time substantially while providing physical interpretability. The effectiveness of PTSTnet derives from three phases (see supplementary materials section 1).

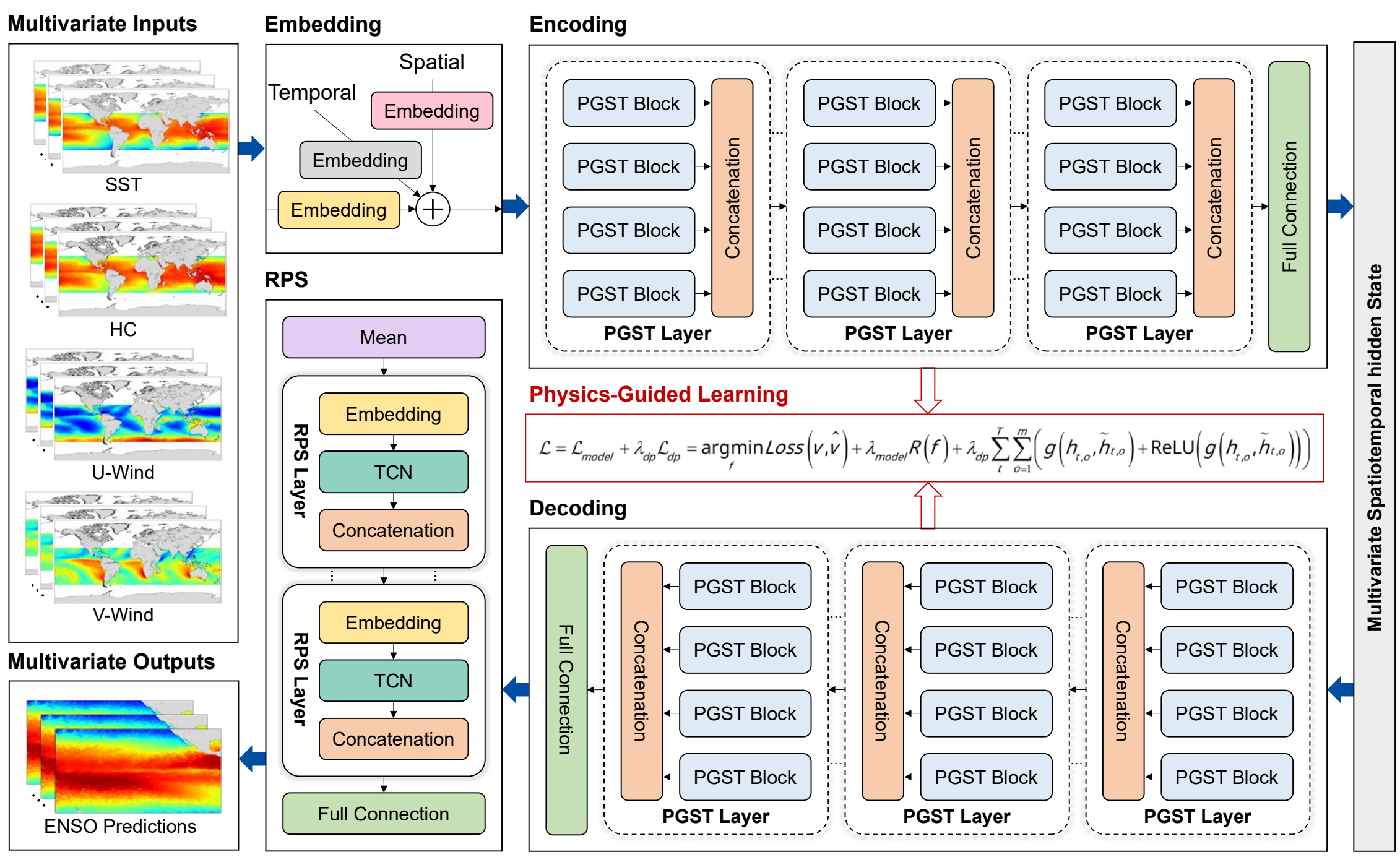

**Fig. 1. Architecture of PTSTnet for long-term ENSO predictions.** Overview of PTSTnet, including the embedding, encoder, decoder, and recurrent prediction strategy (RPS). The encoder and decoder consist of multiple physics-guided spatiotemporal deep learning (PGST) layers. The model input includes the global gridded series of sea surface temperature (SST), upper heat content (HC), sea surface zonal wind (U-Wind), and sea surface meridional wind (V-Wind) for 12 consecutive months, while the prediction context corrects the recurrent predictions with RPS. The model outputs the global gridded series of SST and Niño indices for the next 24 months to provide ENSO predictions.

To address the challenge of scale mixing in long-term spatiotemporal predictions, we proposed a cross-scale spatiotemporal fusion learning strategy that incorporates cross-scale spatiotemporal features related to ENSO dynamics into the formalized modelling process. PTSTnet captures spatiotemporal feature representations that simultaneously express spatial structures and temporal dynamics by performing convolutional tensor transform on the hidden states. Furthermore, it mitigates the exponential growth of model parameters while enhancing the representations of the coupled relationships across spatiotemporal scales.

Focusing on the seamless integration of data and physical processes during modelling, we developed a physics-encoding learning framework to learn ocean dynamical processes from spatiotemporal observations. PTSTnet implements physics-encoding learning for the wave and diffusion processes by updating hidden states and incorporating spatiotemporally varying domain knowledge as loss functions into the deep learning model. It optimizes latent space learning, which effectively enhances prediction skill while providing interpretability.

Considering the large-scale and long-term dependency of ENSO in both spatial and temporal dimensions, we adopted an efficient recurrent prediction strategy (RPS). PTSTnet incorporates previous predictions as prior knowledge for producing long-term predictions by embedding temporal and spatial indices into the feature learning process and leveraging the prediction context to reduce cumulative errors during recurrent prediction. It contributes to identifying seasonal and periodic variations in ENSO predictions and effectively extends prediction lead time.

**Skilful long-term predictions with PTSTnet**

We evaluate the prediction skill and value of PTSTnet against state-of-the-art benchmarks for ENSO prediction. Specifically, dynamical and statistical benchmarks are provided by the International Research Institute for Climate and Society (IRI) (*36*), while deep learning benchmarks include multiple ensemble CNN (*20*), 3D-Geoformer (*23*), ENSO-GTC (*37*), and PhyDNet (*38*) (table S5). Fig. 2 and fig. S4 provide a quantitative evaluation of PTSTnet and all benchmarks for the all-season correlation skill and root mean squared error (RMSE) during 2016–2022.

We highlight that PTSTnet consistently achieves state-of-the-art predictions with high confidence, surpassing statistical, dynamical and deep learning benchmarks (Fig. 2A). For lead times of less than 9 months, the dynamical benchmarks exhibit considerably higher correlation skills, although slightly lower than PTSTnet, while the deep learning benchmarks maintain similar trends. Neither the dynamical nor statistical benchmarks provide effective predictions for lead times of more than 9 months. In particular, CNN and PhyDNet both fail to provide predictions for lead times longer than 18 months. Contrastingly, ENSO-GTC and 3D-Geoformer maintain comparable (slightly lower) performance to PTSTnet throughout the entire prediction period but cannot provide predictions for lead times of up to 20 months or longer (fig. S4). Notably, PTSTnet can predict ENSO for lead times of more than 24 months with correlation skills exceeding 0.5, and the 95% confidence intervals, derived using the bootstrap method remain narrow, indicating robust prediction skill (see supplementary materials section 2).

PTSTnet demonstrates higher correlation skills for the Niño 3.4 index across almost all targeted seasons than CNN (Fig. 2B), which is recognized as a typical deep learning benchmark. The enhancement in correlation skills is particularly significant for targeted seasons between late boreal spring and autumn. Specifically, the predictions targeting the May-June-July (MJJ) seasons show correlation skills exceeding 0.5 only for lead times of up to 11 months in CNN, while extending to 18 months in PTSTnet. This reduction in the prediction gap suggests that PTSTnet is less sensitive to the spring predictability barrier (SPB) (*39*).

We conclude that PTSTnet provides reliable ENSO predictions for lead times of up to 24 months or even longer, which is unattainable with existing state-of-the-art benchmarks (fig. S4). Besides the inherent advantages of the cross-scale spatiotemporal fusion learning against dynamical and statistical models, PTSTnet's performance enhancement is attributed to the successful implementation of the physics-encoding learning for dynamical processes. It incorporates dynamical constraints into the cross-scale spatiotemporal learning, and explores the intrinsic mechanisms of dynamical processes through the physics-encoding spatiotemporal deep learning, effectively reducing data demands for modelling and achieving higher prediction skill with limited training data (fig. S15 and fig. S16). These results demonstrate that PTSTnet leverages the strengths of deep learning for long-term predictions and dynamical models for well-defined physical processes.

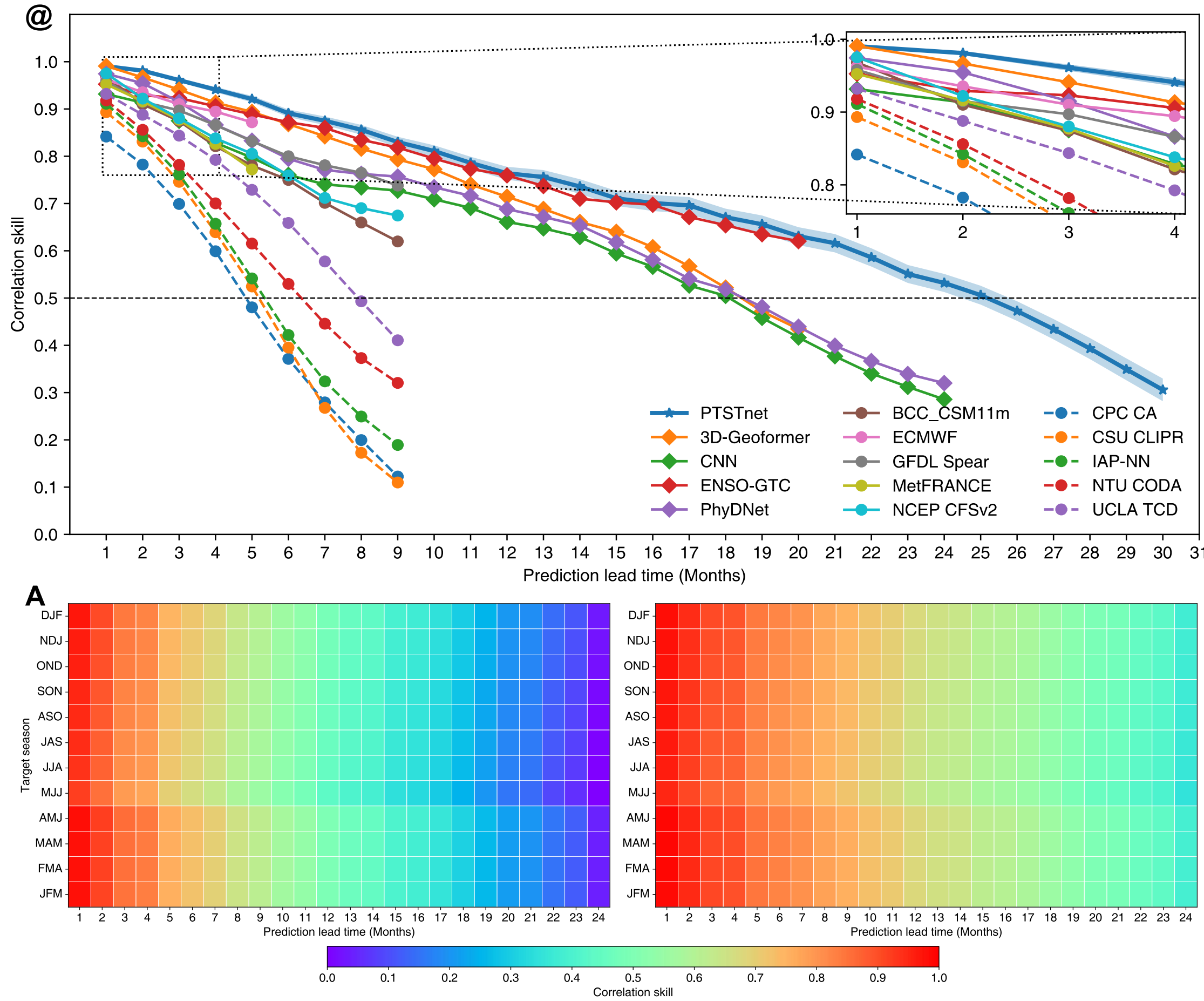

**Fig. 2. Skilful and well-calibrated ENSO predictions by PTSTnet.** (**A**) Comparison of all-season correlation skills between PTSTnet and state-of-the-art benchmarks: PTSTnet (blue line), three deep learning models (solid squares), five dynamical models originating from IRI (solid dots), and five statistical models from IRI (dashed dots). Shading around the PTSTnet line represents 95% confidence intervals based on the bootstrap method. The black dashed line indicates the correlation skill of 0.5. (**B**) Correlation skill of predictions for CNN (left panel) and PTSTnet (right panel) for each calendar month.

## Model interpretability for PTSTnet

While providing skilful ENSO predictions, the intermediate outputs of PTSTnet also offer illuminating insights into the analysis of dynamical processes. To better understand the behaviour of PTSTnet, we conduct interpretable analyses to reveal its ability to learn ENSO dynamics and explain the sources of its interpretability.

### *Interpretable insight into ENSO dynamics*

We visualize the propagation direction of the dynamics learned by PTSTnet at multiple lead times on the test dataset in comparison with the RMSE derived from observations (see supplementary materials section 4). Fig. 3A illustrates that the learned propagation direction aligns well with observations, including the variation in prediction error along the longitudinal and latitudinal directions. This finding confirms that PTSTnet can capture interpretable dynamical propagation processes across different spatiotemporal scales by learning directly from observations. Thus,

PTSTnet serves as a hybrid data-physics-driven model for extracting interpretable insights directly from multiscale ocean dynamics.

The all-season Niño index predictions for lead times of up to 24 months show that PTSTnet accurately predicts the ENSO amplitude (fig. S13 and fig. S14). To reveal the common factors affecting the success of long-term predictions with PTSTnet, we perform a composite analysis of the corresponding ENSO events. Fig. 3B illustrates how spatial error progressively propagates from the western equatorial Pacific to global scales for lead times ranging from 3 to 24 months. Specifically, we observe that the evolution of short-term (6–12 months) and long-term (15–24 months) prediction errors differs in both magnitude and pattern. This evolution occurs in two stages: stage 1 reflects the latitudinal distribution of the prediction error (top panel of Fig. 3B), whereas stage 2 reflects the severe meridional propagation of the prediction error to global scales (bottom panel of Fig. 3B). This finding highlights the physical consistency in the spatiotemporal evolution learned by PTSTnet, especially sea surface temperature (SST) variations along longitudes in the tropical Pacific and its associated atmospheric circulation processes. Especially in the eastern Pacific, the learned propagation process exhibits westward spreading patterns, which aligns with the phenomenon of cold-water upwelling and wind field intensification during ocean–atmosphere interactions (fig. S17). This interaction between the sharp decline in SST and the intensification of trade winds drives ENSO, which enforces the physical consistency learned propagation direction with spatiotemporal observations.

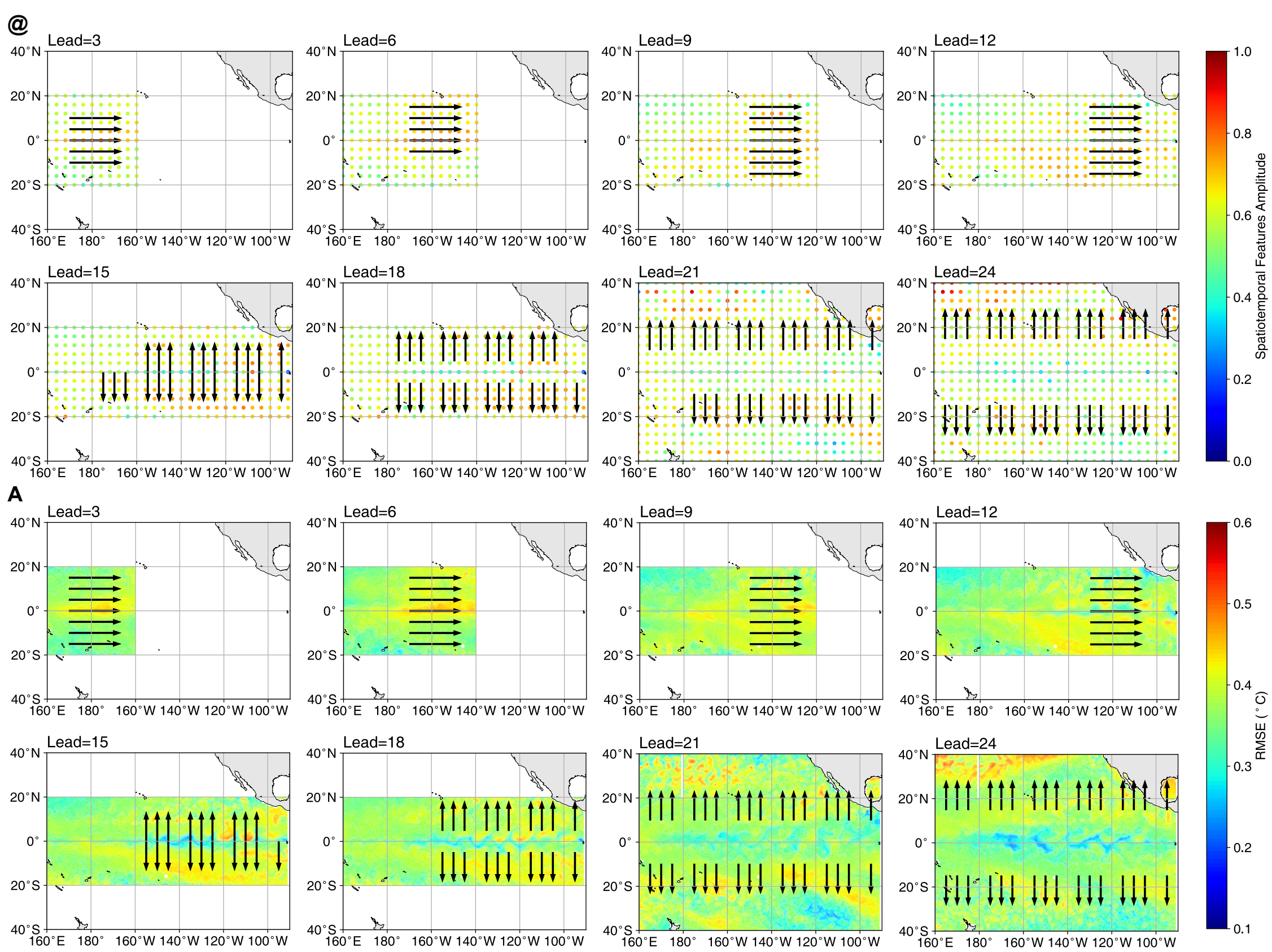

**Fig. 3. Linking the spatiotemporal propagation directions learned by PTSTnet to the dynamical processes presented in observations.** (**A**) Propagation directions of the spatiotemporal dynamics learned by PTSTnet. The black arrows indicate the spatiotemporal dynamics propagation direction. (**B**) Monthly prediction errors of PTSTnet during the test period. The darker regions represent larger prediction errors. The top panels reflect the latitudinal spread of

the prediction errors (stage 1). The bottom panels represent the meridional spread of the error to the global scale (stage 2). The black arrows indicate the error propagation direction.

### *Cross-scale fusion improves interpretability*

The intrinsic properties and fundamental coupling within the ocean–atmosphere system exhibit spatiotemporal scale-dependent relationships, which provide predictable information at various lead times. PTSTnet captures these dependencies through cross-scale spatiotemporal fusion learning. We perform visual comparisons of the spatiotemporal feature maps mined by PTSTnet with different input sequence lengths, to validate the effectiveness and elucidate how learned scale-specific information contributes to ENSO prediction (see supplementary materials section 5). This fusion learning strategy applies a convolutional tensor transform to the hidden states of PTSTnet to learn spatiotemporal features across multiple scales that are highly related to ENSO dynamics, which is consistent with classic ENSO theory. Furthermore, the interpretability of PTSTnet is enhanced through feature selection and scale interaction. These results corroborate that PTSTnet provides an in-depth understanding to reveal the cross-scale spatiotemporal relationships underlying ENSO, thus improving model interpretability.

### *Physical interpretability sources*

We conducted sensitivity experiments to investigate the interpretability sources of PTSTnet from variable, temporal, and spatial perspectives (figs. S6 to 8, fig. S19, and supplementary materials section 6). The enhancement of prediction skill stems not only from cross-scale spatiotemporal fusion learning but also from incorporating multiple oceanic and atmospheric variables that are highly related to ENSO, in contrast to the limited variables of previous research. Notably, PTSTnet incorporates physically reasonable precursors into the formalized modelling of physics-encoding learning by distinguishing precursors across various lead times, thereby enhancing its learning potential for ENSO dynamics.

The successful implementation of PTSTnet simulations highlights the critical role of physics-encoding learning in model configurations. Specifically, ENSO-related dynamical processes are incorporated into the formalized modelling of deep learning, which enables the representation of physics-based model behaviour. The inferred propagation directions of ocean processes underscore the hybrid data-physics-driven paradigm to turn data into insights for oceanographic science. To fully validate the self-interpretation of PTSTnet, we perform interpretability analyses from multiple perspectives, which reveal its learning ability for ENSO dynamical processes and explore its interpretability source. This marks a significant advancement in hybrid data-physics-driven modelling for ENSO prediction.

## Accurate predictions of ENSO events

### *Skilful prediction for super El Niño*

To elucidate the spatiotemporal covarying dynamics more explicitly, we highlight PTSTnet's superiority in long-term ENSO predictions using the 2015–2016 super El Niño event as a case study, complemented by quantitative statistical results. The 2015–2016 El Niño, one of the strongest events on record, remains challenging to predict using dynamical models initialized from spring 2015 due to large uncertainty (*40*). Consequently, this ENSO event is appropriate for evaluating the effectiveness of PTSTnet. Our model accurately describes the spatiotemporal evolution of ocean temperatures in the tropical Pacific.

PTSTnet accurately simulates fundamental observed characteristics of ENSO (Fig. 4). The temporal evolution of the Niño 3.4 predictions closely aligns with observations, albeit with slightly larger magnitudes (Fig. 4A). In addition, PTSTnet realistically simulates the observed spatiotemporal evolution of upper-ocean temperature anomalies (Fig. 4B). ccurate spatiotemporal modelling enhances prediction skill: PTSTnet can characterize the locations and magnitudes of SST anomalies for lead times of more than 24 months and distinguish different types of El Niño events for lead times of up to 12 months (fig. S9). Furthermore, PTSTnet also accurately reproduces the spatiotemporal coupling relationships, including the synchronizations and autocorrelations between SST anomalies and sea surface wind (SSW) (fig. S24 and fig. S25). Simulating these observed relationships further highlights PTSTnet's effectiveness in predicting multivariate synergistic dynamics across various spatiotemporal scales.

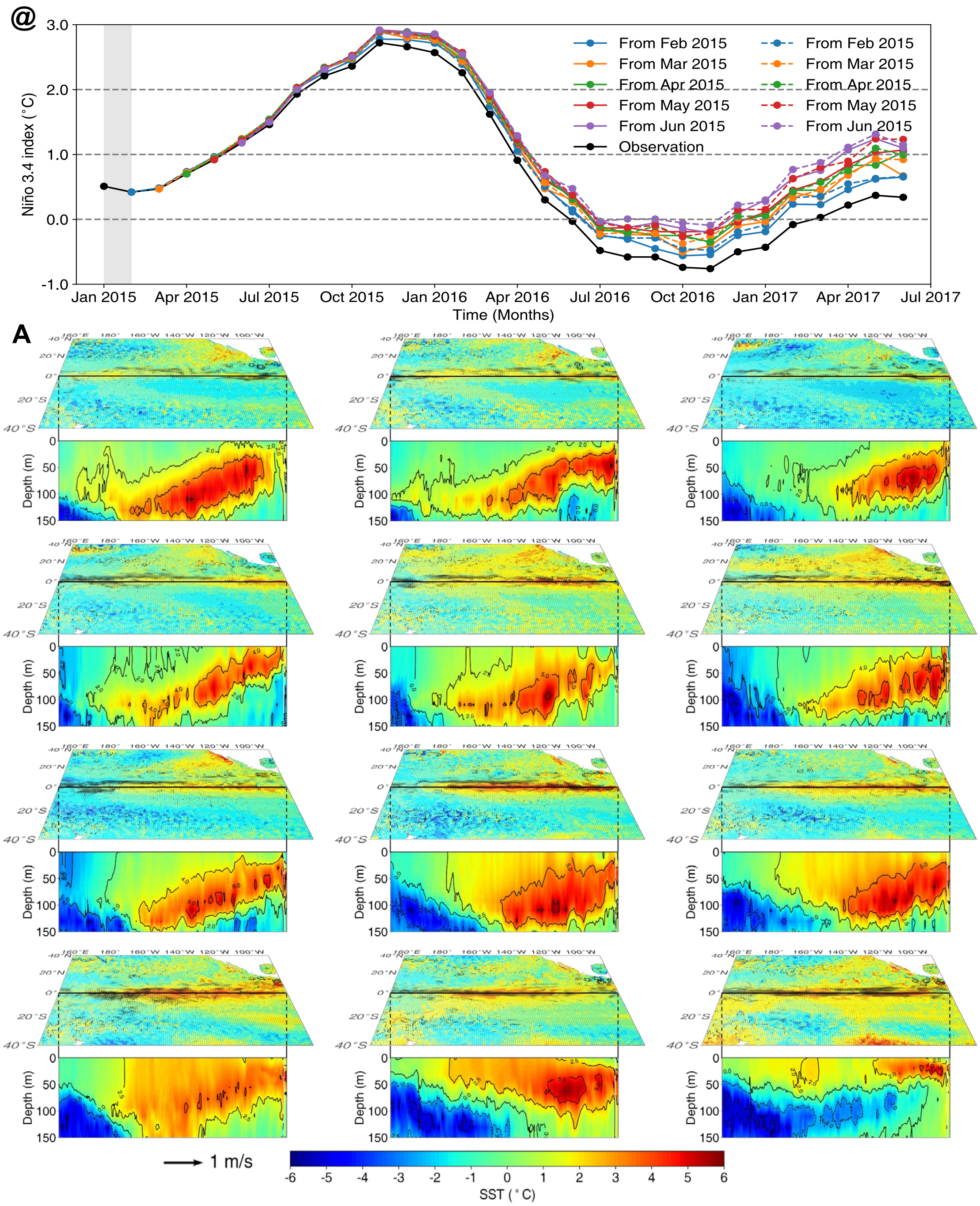

**Fig. 4. PTSTnet-predicted Niño 3.4 SST anomalies and spatiotemporal evolution for the 2015–2016 super El Niño event.** (**A**) Observed (black line) and PTSTnet-predicted (coloured lines) Niño 3.4 SST anomalies, with

predictions initiated from February to June 2015 onward. Two experimental configurations are compared: Solid coloured lines represent predictions using all input predictors including SST, U-Wind, V-Wind, and HC, while dotted coloured lines represent predictions using input predictors including SST and HC only with wind stress anomaly effects removed. (**B**) Twelve-month predicted spatiotemporal evolution initiated from April 2015: SSW (vector) and SST (shading) alongside synergistic upper-ocean temperature anomalies (shading and contours) in the vertical profile along the equator.

### *Multivariate synergies enhance predictability*

The effective and accurate predictions performed with PTSTnet are partially attributed to its appropriate learning of dynamical systems with the physics-encoding learning, which is consistent with the Bjerknes feedback. Specifically, during El Niño, large positive SST anomalies in the eastern equatorial Pacific and covarying positive subsurface temperature anomalies are accompanied by westerly wind anomalies over the central equatorial Pacific around the dateline. PTSTnet predictions for lead times of up to 12 months, initiated from April 2015, clearly reproduce the spatiotemporal evolution of upper-ocean temperature and its relationship with SSW (Fig. 4B, fig. S24, and fig. S25). These multivariate synergies effectively represent the Bjerknes feedback and are comparable with state-of-the-art dynamical models (*41*). Conclusively, PTSTnet can track large-scale coupled ocean–atmosphere variations and successfully predict ENSO with high intensity and long lead times, which are challenging for state-of-the-art models. Moreover, PTSTnet can provide predictions early in the calendar year with relatively low errors, indicating its ability to mitigate or eliminate the negative effects of the SPB (fig. S19).

Overall, PTSTnet provides robust long-term ENSO predictions in all phases, including the super El Niño, multiyear La Niña, and neutral phases, and the predictions align with observations in both magnitude and trend (Fig. 4 and figs. S21 to 28). This is primarily attributed that PTSTnet comprehensively learns the complex variations implicit in ocean and atmosphere dynamics through physics-encoding learning while employing cross-scale spatiotemporal fusion learning to deeply mine information from spatiotemporal observations. Thus, with this sophisticated structure, PTSTnet can simulate ocean–atmosphere energy exchanges almost simultaneously. Moreover, the geoscientific fluid programming in dynamical models typically employs interval flux exchanges and parametric approximations of unknown mechanisms, hindering continuous interactions among various processes. Whereas deep learning models overly rely on the ergodicity of modelling data, leading to limited prediction performance for extreme events with low probability. Therefore, PTSTnet with the hybrid data-physics-driven paradigm achieves superior performance in various scenarios compared with state-of-the-art benchmarks.

## Discussion

Nowadays, real-time ENSO prediction mainly relies on physics-driven dynamical models with significant bias and uncertainty, thereby limiting the reliability of long-term predictions. Recent advances in data-driven deep learning have provided promising directions for nonlinear system modelling and multiyear ENSO predictions. However, limited by the inherent weaknesses in model scale and non-interpretability of deep learning approaches, accurate predictions of long-term ENSO evolution remain challenging.

Inspired by recent successful applications of the sought-after physics-guided deep learning in computational fluid dynamics, we present PTSTnet, a physics-guided tensor-train spatiotemporal deep learning model for ENSO prediction. Our model integrates cross-scale spatiotemporal fusion learning and recurrent prediction strategy within an innovative physics-encoding learning framework. To the best of our knowledge, this represents the first implementation of hybrid modelling with deep learning and physical processes for ocean prediction. The inherent advantage of physics-encoding learning in PTSTnet enhances long-term prediction skill and reduces data demands by combining deep learning with cross-scale dynamical processes in the spatiotemporal domain, thereby establishing the multivariate synergies of ENSO dynamics (fig. S15). PTSTnet demonstrates its effectiveness and preeminence in ENSO prediction, as it accurately simulates the coupled interactions between upper-ocean temperature anomalies and SSW, outperforming state-of-the-art benchmarks.

Specifically, PTSTnet can predict ocean temperature anomalies in the tropical Pacific at lead times of more than 24 months, as demonstrated by Niño index predictions for the last two years. Notably, the predictions for various ENSO events highlight the impressive performance of PTSTnet in characterizing multiscale ocean–atmosphere interactions. Furthermore, the physical interpretability of PTSTnet is validated through the error propagation of dynamical processes and the contributions of cross-scale information in modelling. The sensitivity experiments further confirm the sources of PTSTnet's interpretability and emphasize the effectiveness of physics-encoding learning in representing dynamical processes, thereby mitigating the SPB.

The intrinsic opacity of deep learning poses a fundamental challenge to the interpretability of ENSO predictions (*10*, *23*, *26*). PTSTnet can predict ENSO evolution at lead times of more than 24 months while also providing corresponding physical interpretability. We analyze the error propagation within the spatiotemporal multiscale dynamics produced by PTSTnet to emphasize the advantages of this hybrid data-physics-driven model in capturing typical precursors in ENSO prediction, which is further corroborated through sensitivity experiments. Moreover, the cross-scale spatiotemporal dynamics learned by PTSTnet are consistent with classic ENSO theory, indicating that our model can comprehensively explore and explicitly capture the physical processes underlying ENSO complexity. This is the most significant motivation that distinguishes PTSTnet from existing deep learning models. The design and evaluation of interpretable sensitivity experiments are crucial for deepening our understanding of ENSO dynamics and validating PTSTnet's reliability. Thus, PTSTnet can provide not only interpretable representations of ENSO dynamics but also insights into fundamental contributions affecting long-term ENSO predictions.

Further improvements are necessary in certain aspects of PTSTnet's configuration. Specifically, given that ENSO is governed by the ocean–atmosphere interactions involving multiple dynamical processes (*6*, *10*, *17*). PTSTnet could benefit from incorporating additional processes, such as momentum conservation, through physics-encoding learning to enhance its prediction skill. However, integrating higher-order dynamical processes into deep learning models for long-term spatiotemporal prediction remains a significant challenge and constitutes the primary focus of our future work. Additionally, selecting appropriate transfer learning approaches for specific application scenarios remains one of the potential ways to enhance model performance. Transfer learning-based algorithms that freeze shallow parameters during fine-tuning (*20*, *25*, *42*), combined with optimized loss functions to address skilful prediction under data scarcity, deserve further exploration in formalized modelling.

The successful implementation and robust performance of PTSTnet in ENSO prediction underscore its unique potential to address increasingly complex geophysical challenges. This lays the foundation for the broader application of hybrid data-physics-driven models in various ocean modelling tasks and is promising to catalyze revolutionary paradigm shifts in geoscience.